# van der Waals Nanoreactors

Zhaoyi Joy Zheng[1,2,#], Haosen Guan[1,#], Danrui Ni[3], Guangming Cheng[4], Yanyu Jia[1], Ipsita Das[1], Yue Tang[1], Ayelet July Uzan-Narovlansky[1], Lihan Shi[1,2], Kenji Watanabe[5], Takashi Taniguchi[6], Nan Yao[4], Robert J Cava[3], Sanfeng Wu[1*]

[1] Department of Physics, Princeton University, Princeton, New Jersey 08544, USA
[2] Department of Electrical and Computer Engineering, Princeton University, Princeton, New Jersey 08544, USA
[3] Department of Chemistry, Princeton University, Princeton, New Jersey 08544, USA
[4] Princeton Materials Institute, Princeton University, Princeton, New Jersey 08544, USA
[5] Research Center for Electronic and Optical Materials, National Institute for Materials Science, 1-1 Namiki, Tsukuba 305-0044, Japan
[6] Research Center for Materials Nanoarchitectonics, National Institute for Materials Science, 1-1 Namiki, Tsukuba 305-0044, Japan

[#] These authors contributed equally to this work
[*] Email: sanfengw@princeton.edu

**Abstract**

**Advancing the chemical synthesis of crystals is important for both fundamental research and practical applications of quantum materials. While established bulk-phase and thin-film growth methods have enabled enormous progress[1–4], synthesizing single crystals suitable for quantum electronic discoveries remains challenging for many emerging materials. Here, we introduce van der Waals (vdW) stacks[5–7] as nanochemical reactors for single-crystal synthesis and demonstrate their broad applicability in growing both elemental and compound crystals at the micrometer scale. By encapsulating atomically thin reactants that are stacked compactly with inert vdW layers such as hexagonal boron nitride (hBN)[8], we achieve nanoconfined synthesis with the resulting crystals remaining encapsulated. As proof of concept, we synthesized isolated single crystals of elemental tellurium and distinct types of Pd-Te compounds. Structural characterization, including atomic-resolution scanning transmission electron microscopy, confirms the high crystalline quality of the products. We confirm the intrinsic semiconducting gap of tellurium and observe that non-stoichiometric PdTe$_{1-x}$ with a significantly reduced Te content (x ≈ 0.18, a regime not previously achieved) retains uniform crystallinity and exhibits superconductivity below a critical temperature of 3.8 K. This nanochemical synthesis is broadly generalizable, chip-integrable, well-suited to a wide range of processing conditions, and compatible with nanofabrication routines for constructing devices. The concept of vdW nanoreactors offers a powerful and versatile pathway to expand the accessible landscape of quantum materials.**

**Main**

Chemical reactions in nanoconfined spaces isolated from the surroundings can be fundamentally different from those in open environments[9,10]. Controlling and understanding chemical reactivity and material transformations in nano- and micro-scale reactors are essential to addressing challenges across multiple disciplines spanning from life science and biotechnology to catalysis and nanotechnology.

Various nanoreactors have been explored for investigating the formation of nanoparticles and nanocrystals under confinement in different context[9–11], using a range of architectures including rigid nanotubes[12–14], polymer assemblies[9,15], proteins[16,17], mesoporous templates[18,19], microfluidic droplets[20–22] and lithographically defined patterns[23,24].

Exploring nanoreactors for synthesizing quantum materials is to date a largely unexplored arena. Mature synthetic routes to inorganic crystalline materials for investigating quantum electronic phenomena are bulk-phase and thin-film growth in solid-state, solution, or vapor-based open chemical environments[1–4], leaving behind vast opportunities to be explored in the small scale under nanoconfinement. To advance, nanoreactors capable of accelerating quantum materials discoveries would require several key features simultaneously that are however mostly not available in previously developed nanoreactors[9,10]: (1) The final product is in a single-crystal form with a micrometer size at least in one direction (i.e., not merely nanoparticles); (2) The reaction is generalizable to various combinations of reactants with controlled stoichiometric compositions for synthesizing a broad class of materials; (3) The nanoreactors can support harsh synthetic conditions, such as high temperatures or high pressures, often necessary for quantum materials synthesis; (4) The product crystal is isolated from unwanted substances; (5) The synthesis is compatible with further nanofabrication processes for constructing devices from the as-grown crystals, essential for investigating its quantum electronic or photonics properties. In this work, we introduce the concept of van der Waals (vdW) nanoreactors (**Fig. 1**), which feature all aspects mentioned above.

**vdW nanoreactors**

vdW stacks of atomically thin 2D crystals have been one of the foci in condensed matter research since it enables the integration of distinct materials and physical phenomena at the atomic scale with striking modification of low energy electronic structures through, e.g., moiré quantum engineering[5–7,25–29]. Emphasis has been so far on the modulation of the physical properties, instead of chemical reactivity, of the materials encapsulated in vdW stacks. Here we change the mindset to view vdW stacks as an extraordinary mixture of atomically thin reactants that are packed closely together at the atomic scale (**Fig. 1a**), unprecedented to conventional solid-state grinding methods used in bulk growth. An inert layer, such as hexagonal boron nitride (hBN), can be used to encapsulate and isolate the mixture from the environment, forming a confined nanoreactor (**Fig. 1b**), which allows for high temperature treatment [30] and prevents the loss of masses to the environment or contamination. **Fig. 1a** illustrates the flowchart for creating vdW nanoreactors. Atomically thin layered reactants are first exfoliated onto a substrate (e.g., the typical $SiO_2$/Si), which may or may not undergo further processes (such as oxidization, etc) to modify its chemical compositions. A series of precursor layers, together with the encapsulation layers, can then be picked up subsequently and stacked to form a vdW stack using the standard transfer techniques used in 2D material community. Chemical reactions will be triggered inside the stack by methods such as heating (used in this work) and/or pressure (**Figs. 1b & c**). The hBN encapsulation layer can support reactions up to ~ 1000 °C[30] (**Extended Data Fig. 1**), allowing for its use in the synthesis of a wide class of materials. We note that due to nanoconfinement, reactions between the atomically thin and compact reactants within the vdW stacks can occur at temperatures significantly lower than that required for conventional crystal growth.

We highlight two important aspects associated with the vdW nanoreactors. First, the choices of precursor layers are immensely diverse. It has been estimated that more than 1,000 layered crystals can be exfoliated down to the 2D limit[31–33], each of which may be further chemically treated to form derived structures. All these can potentially be used as reactants. Thin film depositions can also introduce non-layered materials, such a thin layer of metal, to be included. The ability to control atomic composition and stoichiometry in the vdW nanoreactors is hence exceptional. Second, techniques for characterizing the final crystals, including both atomic structures and its electronic and optical properties, are readily available thanks to the developments for 2D materials. For instance, crystals formed inside the vdW stacks can readily be turned into electrical transport devices using established 2D nanofabrication methods. Below we demonstrate the methodology and the applications of vdW nanoreactors by achieving the growth and characterization of distinct representative crystals.

**vdW nanoreactors for synthesizing tellurium single crystals**

We begin by illustrating the creation of a vdW nanoreactor consisting of a hBN/MoTe$_2$/oxidized MoTe$_2$/hBN stack where the reactants are intrinsic few layer 2H-MoTe$_2$ and its oxide (**Fig. 2a**). The oxidization of MoTe$_2$ on the hBN bottom layer is achieved by thermal annealing at 280 °C in an oxygen environment, resulting in an oxygen-rich amorphous Mo-Te-O mixture that is consistent with the formation of MoO$_3$ and TeO$_x$ (**Extended Data Fig. 2**). The fully encapsulated stack is then heat-treated at 350°C for about 30 minutes. An optical image of a stack (Growth T1) after reactions is shown in **Fig. 2b**. A bubble is formed between the two hBN layers, likely caused by the vapor pressure developed due to e.g., sublimation of the oxides (such as MoO$_3$) inside the vdW stack at high temperatures. Inside or near the bubble, wire-like new materials forms, which we confirm to be tellurium single crystalline wires. The process is highly reproducible, as shown in **Figs. 2c & d** for another two separate growths (T2 & T3), all yielding similar results. Scanning electron microscope (SEM) images shown at the bottom panels of **Figs. 2c & d**, reveal clearly the long rectangular morphology expected for the quasi-1D crystal structure of tellurium (**Figs. 2e & f**).

To confirm the high crystallinity of the product, we perform atomic-resolution scanning transmission electron (S/TEM) microscope studies on the vdW stack after reaction (see **Extended Data Fig. 3** for the fabrication process of the TEM samples). The results are shown in **Figs. 2g – i**, where the Te atoms in the wire are clearly revealed, confirming the crystal structure consistent with that of tellurium single crystal (**Fig. 2f**). No single defect appears in the observation (**Fig. 2i**). We further perform elemental mapping using energy-dispersive X-ray spectroscopy (EDX) around a selected wire, as shown in **Figs. 2j**, where Mo and O atoms are absent at the location of Te crystal while they do form substances nearby. The essential reaction process in this vdW nanoreactors may be approximately described as MoTe$_2$ + MoO$_3$ → Te + MoO$_2$ + …. We emphasize that such process occurs only because of the nanoconfinement by the hBN layer, without which all vaporized substances at high temperatures will disseminate into the open environment (**Extended Data Fig. 1**).

Elemental tellurium crystals have recently attracted substantial interests due to its potential applications in electronics[34–38], thermoelectricity[39], chiral materials[37,38], and nanoprosthesis[40]. Intrinsic tellurium is a semiconductor with a bandgap of ~ 340 meV.[41] However recent experiments using solution-based growth typically yield a slight hole doping, presumably attributed to Te vacancies.[37,38] To examine the electrical transport properties of the Te wire created in our nanoreactors, we deposited metal contacts onto the wire followed by etching the hBN layer only at the contact region (the non-contact region of Te is still protected by hBN), employing typical electron beam nanolithography processes. The SEM image of a device is shown as inset in **Fig. 2k**. We observe activated transport behavior in the resistance–temperature dependence of as-grown tellurium. The Arrhenius fits yield bandgaps of approximately 320 meV and 280 meV from two distinct contact regions (Fig. 2k), both close to the intrinsic value. The transport data indicates substantially reduced vacancies or impurities in the sample. The observation not only confirms the semiconducting nature of our as-grown tellurium, but also implies excellent quality of the crystals obtained in our nanoreactor approach.

**vdW nanoreactors for synthesizing Pd-Te binary crystals**

To demonstrate the versatility of the vdW nanoreactors, we now modify the previous reaction by introducing Pd into the reaction, such that the reaction may be engineered towards producing distinct crystals, with targets of Pd-Te binary compounds in mind. We introduce Pd by simply replacing $MoTe_2$ with $Pd_7MoTe_2$ **(Fig. 3a)**, with a hypothetical synthetic path of $Pd_7MoTe_2 + MoO_3 \rightarrow Pd_xTe_y + MoO_2$ + ….The creation of atomically thin layer of $Pd_7MoTe_2$ follows the procedure developed in previous works[42–44], which discovered that 2D metal of Pd forms upon heat treatment when a Pd source is in contact with 2D $MoTe_2$ layers. The 2D Pd is precisely 7 atomic layers per $MoTe_2$ layer, forming a new compound of $Pd_7MoTe_2$. To form a vdW nanoreactor, we transfer a layer of $Pd_7MoTe_2$ onto oxidized $MoTe_2$ as reactants, fully encapsulated by top and bottom hBN. We then heat the stack to about 400 ºC in a typical thermal annealing process (see Methods), after which the result is examined under microscopes.

Instead of wires we now find crystals of very different morphologies (often elongated hexagons), as shown in **Figs. 3b-d** for three different samples (Growth PT1-3), already implying that a different type of crystal is formed. STEM plan-view examination of location 1 in Growth PT1 (indicated by the blue square in **Fig. 3c**) after reactions is shown in **Figs. 3e-j**, where two products are clearly identified. The brightest area with sharp edges seen in the high-angle annular dark-field (HAADF) image (**Fig. 3e**) corresponds to a binary crystal formed by Pd and Te, as revealed by EDX elemental mapping (**Figs. 3f**). There is $MoO_2$ formation nearby as seen clearly in the Mo and O maps. More EDX analysis can be found in **Extended Data Fig. 4**. Atomic resolution STEM images show that the Pd-Te compound formed here develops an excellent crystallinity, where the atomic structure can be identified as that of $Pd_9Te_4$ (**Figs. 3g-j**).

It is known that the Pd-Te compounds exhibit a variety of phases with different stoichiometric compositions.[45,46] In addition to the formation of $Pd_9Te_4$ crystals in this nanoreactor (Growth PT1), we have indeed also observed other binary species. As highlighted in location 2 of Growth PT1 (indicated by the purple square in **Fig. 3c)**, there is formation of another phase in this vdW nanoreactors. Its structural characterization is shown in **Figs. 3k-q**, where EDX mapping confirms the binary composition with an atomic ratio of Pd:Te to be about 55%:45%. However, $Pd_{11}Te_9$ doesn't exist as a standalone stoichiometric phase in the Pd-Te phase diagram [45,46]. Atomic resolution images (**Figs. 3m-p**) reveal that the lattice structure closely resembles that of PdTe (similar to the hexagonal NiAs structure with space group $P6_3/mmc$). We hence attribute this phase as $PdTe_{1-x}$ with a substantially reduced content of Te (x ≈ 0.18). We are not aware of other reports that produce this much reduction of Te. We note that the lattice of PdTe could retain its form likely thanks to the relatively low growth temperature used in our reaction and that the Pd richness is reasonable in our nanoreactors because Pd is significantly more in the reactants loaded initially. In the next section, we reproduce the growth of this $PdTe_{0.82}$ compound and perform careful transport characterization.

**Superconductivity in $PdTe_{0.82}$**

In Growth PT4, we reproduced the growth of non-stoichiometric PdTe$_{1-x}$ (x ≈ 0.18) that is large enough for creating transport devices as seen in **Fig. 4a,** which shows a false-colored SEM image of the device after depositing gold electrodes. We first present data confirming the high-quality crystal structure using cross-section STEM performed on the same device after transport measurement. (**Figs. 4b-d**). Followed by a cut of the device along the black dashed line shown in **Fig. 4a** using a focused ion beam, we verify again that the atomic composition to be ~ 55%:45% (**Fig. 4b**). Atomic resolution STEM images confirm that the entire compound is a uniform single crystal, with an indistinguishable lattice at different locations (**Figs. 4d** & **Extended Data. Fig. 5**). This lattice is again consistent with that of PdTe viewed from the cross-section angle shown in **Fig. 4c**. We hence confirm that the same PdTe$_{1-x}$ with x ≈ 0.18 is selected.

We present quantum transport properties of this PdTe$_{0.82}$ crystal with a measurement configuration shown in **Fig. 4a**. The crystal displays a metallic behavior with a four-probe resistance $R_{xx}$ on the order of 1 Ω at room temperature. Upon cooling, the resistance slowly decreases and suddenly drops to zero below a critical temperature, $T_c$, of ~ 3.8 K (**Fig. 4e**), signifying the formation of superconductivity. **Fig. 4f** plots the $IV$ curves taken at different temperature ($T$) below and above the transition, revealing the typical nonlinear characteristics expected for a superconductor. Differential resistance d$V$/d$I$ (**Figs. 4g** & **h**) highlights the critical current of ~ 130 uA at low $T$, a value that decreases upon warming up the sample. The application of an external magnetic fields, $B\perp$, perpendicular to the plane, suppresses the superconductivity (**Fig. i**) above a critical field of ~ 380 mT at low $T$. We extract the critical magnetic field, $B_{c,\perp}$, as a function of temperature (**Fig. 4j**), which follows the linear Ginzburg-Landau (GL) form $B_{c,\perp} = \Phi_0/(2\pi\xi_{GL}^2)(1-T/T_c)$ near $T_c$, where $\xi_{GL}$ is the extrapolated GL coherence length at zero temperature and $\Phi_0$ is the flux quantum. We find $\xi_{GL}$ to be approximately 27 nm. Superconductivity has been found in various Pd-Te compounds with distinct $T_c$, with the highest known to date as that of PdTe ($T_c$ ~ 4.5 K)[47–49] and all other known phases exhibit a much lower $T_c$. Our observation of a slighter lower $T_c$ (3.8 K) and a much higher $B_{c,\perp}$ compared to PdTe[47–49] is consistent with the structural characterization of it being a PdTe-type crystal but with a much lower Te concentration. It also demonstrates that our approach can achieve significant variations of atomic contents in crystals. Superconductivity is observed over the entire crystal, as implied by data taken from another pair of contacts (**Extended Data Fig. 6**).

**Discussion and outlook**

Our work establishes a new synthetic pathway to high quality crystals though direct on-chip chemical processes under nanoconfinement employing a novel concept of vdW nanoreactors. The two reactions demonstrated here represent only the initial examples of what vdW nanoreactors can achieve. We anticipate that many more quantum materials will soon be synthesized and characterized using this approach. A vast array of ultrathin materials — whether exfoliated[31–33], deposited, chemically modified, or fabricated by other means — can be stacked to form vdW nanoreactors. Within these structures, reactions can be initiated not only by heating but also through alternative stimuli such as high pressure or laser irradiation, opening an expansive space for exploration.

One particularly promising direction is the search for improved superconductors and topological quantum materials that are challenging to obtain via conventional bulk synthesis. There are at least two key advantages to using vdW nanoreactors in the context of superconductivity. First, the resulting materials are microscale single crystals compatible with nanofabrication, naturally minimizing issues like fractional superconducting volume often encountered in bulk samples in the search of new superconductors. Second, chemical reactivity under nanoconfined conditions differs significantly from that in open environments, offering access to materials with composition and stoichiometry that may be difficult to achieve using traditional methods but important to optimizing superconductivity. Another exciting avenue is leveraging vdW nanoreactors to deepen our understanding of nanoscale chemistry and molecular dynamics under confinement — a topic of broad relevance to fundamental research at small scales. By carefully selecting materials and reaction conditions, these systems can serve as well-defined platforms for studying chemical processes and material transformations in nanoscale. The resulting insights could inform diverse fields, including biological systems, nanofluid channels, atomic-scale manufacturing, and emerging small-scale technologies.

## Acknowledgement

This work is supported by AFOSR (on device fabrications and transport) through awards FA9550-23-1-0140 and FA9550-25-1-0354 to S.W. and NSF (on chemistry and materials characterizations) through the Materials Research Science and Engineering Center (MRSEC) program of the National Science Foundation (DMR-2011750) awarded to S.W., R.J.C., and N.Y., and the MRSEC's partial support to Princeton's Imaging and Analysis Center (IAC). S.W. acknowledges the support from the Gordon and Betty Moore Foundation through Grant GBMF11946. K.W. and T.T. acknowledge support from the JSPS KAKENHI (Grant Numbers 21H05233 and 23H02052), the CREST (JPMJCR24A5), JST and World Premier International Research Center Initiative (WPI), MEXT, Japan.

## Author Contributions

S.W. conceived and designed the project. Z.J.Z and H.G. fabricated the nanoreactors, performed synthesis, conducted experiments characterizing the synthesized materials (including both structural and electronic properties) and analyzed the data, assisted by Y.J., I.D., Y.T., A. J. U.-N., and L.S., and supervised by S.W. D. N. and R.J.C grew and characterized bulk $MoTe_2$ crystals. K. W. and T. T. provided hBN crystals. Z.J.Z, Y.J., G.C., and N.Y. performed TEM measurements. S.W., Z.J.Z., H.G., D.N., R.J.C, G.C., and N.Y. analyzed the data and interpreted the results. S.W. and Z.J.Z wrote the paper with input from all authors.

## Competing interests

The authors declare no competing interests.

## Data Availability

The data that support the findings of this study are available from the corresponding author upon reasonable request.

## Code Availability
N.A.

## Methods

### Fabrication of vdW nanoreactors for Te growth

To prepare the vdW nanoreactors, $MoTe_2$ flakes and hBN were mechanically exfoliated on $SiO_2/Si$ substrates. We chose 2 ~ 4 layer $MoTe_2$ which were identified under an optical microscope located in an Ar-filled glovebox. $MoTe_2$ flakes were then placed on top of a 10-15 nm thick hBN flake to form a $MoTe_2$/hBN stack on a $SiO_2/Si$ substrate using standard 2D dry transfer techniques. The stack was subsequently placed in a furnace and treated at 280 °C for 1 hour with $O_2$ gas flow at a rate of 0.1 L/min. After the $MoTe_2$ oxidization process, another hBN flake (15-20 nm thick) and a fresh $MoTe_2$ (2 ~3 layer) flake was transferred onto this Mo-Te-O oxide/hBN bottom stack. To trigger the reaction and induce Te growth, the final stack (hBN/$MoTe_2$/Mo-Te-O oxide/hBN) was placed in the furnace and heated up to 350 °C for 30 minutes, with oxygen at a flow rate of 0.1 L/min. Growth T1-3 all followed the above process.

### Fabrication of vdW nanoreactors for Pd-Te crystal growth

To prepare the vdW nanoreactors for Pd-Te crystal growth, the bottom oxidized $MoTe_2$/hBN stack was prepared using the same method as described in the Te growth. To prepare the $Pd_7MoTe_2$ reactant, we stack another hBN flake and a bilayer $MoTe_2$ together and subsequently placed them on a pre-deposited thin Pd source on a $SiO_2$/Si substrate. The Pd source was prepared using standard electron beam lithography, followed by cold development, reactive ion etching, and metal deposition to deposit the patterned Pd sources of about 20 nm thickness on $SiO_2$/Si wafers. This hBN/$MoTe_2$/Pd stack was annealed in a furnace at 300 °C for 1 hour under a forming gas flow (95% $N_2$ + 5% $H_2$) at 0.1 L/min to induce 2D Pd diffusion, resulting in the formation of a $Pd_7MoTe_2$ thin film.[42-44] The resulting hBN/$Pd_7MoTe_2$ stack was then picked up and transferred onto the previously prepared oxidized $MoTe_2$/hBN stack, yielding the final vdW reactor structure: hBN/$Pd_7MoTe_2$/Mo-Te-O oxide/hBN. This final stack was subjected to crystal growth conditions under high temperature treatment up to 400 °C. For Growth PT1, the stack underwent two growth stages: an initial treatment at 350 °C in oxygen (0.1 L/min) for 30 minutes, followed by a second stage at 400 °C in oxygen (0.1 L/min) for another 30 minutes. Growths PT2 and PT3 both involved a first stage at 300 °C in oxygen (0.1 L/min) for 1 hour, followed by a second stage at 400 °C in forming gas (0.1 L/min) for 1 hour. Growth PT4 followed a similar two-step procedure, with the first stage at 300 °C in oxygen (0.1 L/min) for 45 minutes, and the second stage at 400 °C in forming gas (0.1 L/min) for 1 hour.

**Plan-view TEM Sample Preparation**

The plan-view TEM samples were prepared by transferring hBN-encapsulated (from both top and bottom) samples after final growth to TEM grids with holey silicon nitride, using the standard 2D dry transfer techniques employing polycarbonate film. The polycarbonate film left on the stack after the transfer was dissolved in chloroform for 30 min. A cartoon illustration of the sample preparation process can be found in **Extended Data Fig. 3a**.

**Cross-section TEM Sample Preparation**

The cross-section TEM sample were prepared by focused-ion-beam (FIB) cutting the $PdTe_{1-x}$ device (see below on the transport device fabrication) using Helios NanoLab G3 UC dual-beam FIB-SEM. A protective Carbon layer was first deposited using a gas injection system to prevent beam-induced damage during milling. The region of interest was then milled using $Ga^+$ ion beam and extracted using the in-situ lift-out method. The lamella was then mounted onto a TEM grid for imaging. Final thinning and polishing were performed using 2kV $Ga^+$ ion beam till the sample is electron-transparent with a thickness less than 100 nm. A cartoon illustration of the sample preparation process can be found in **Extended Data Fig. 3b**.

**SEM**

The SEM experiments were performed using Verios 460 XHR SEM. The samples were mounted on conductive carbon tape and imaged without additional coating. SEM images were acquired under high vacuum using a secondary electron detector with an accelerating voltage of 5 kV, a beam current of 25 pA, and a working distance of 4 mm.

**S/TEM and EDX**

Atomic-resolution HAADF imaging and EDX mapping were performed using Titan Cubed Themis 300 double Cs-corrected S/TEM equipped with an extreme field emission gun source and a super-X EDS system. The system was operated at 300kV.

**Transport Device Fabrication**

To prepare transport devices for Te nanowires, the hBN encapsulated final stacks after growth were picked up and transferred onto $SiO_2$/Si wafers with prepatterned metal alignment markers. We used electron beam lithography to pattern metal electrodes, followed by cold development. Then we used reactive ion etching to etch through top hBN and used e-beam metal deposition to deposit metal contacts (~5 nm Ti / ~120 nm Au). For the Pd-Te device, the hBN encapsulated final stacks after growth were picked up and transferred onto $SiO_2$/Si wafers with prepatterned metal alignment markers. The top hBN was etched through using reactive ion etching to expose the as-grown crystals. We then employed standard electron beam lithography, followed by cold development, reactive ion etching, and metal deposition to pattern and deposit metal contacts (~15 nm Ti / ~135 nm Au).

**Transport Measurements**

The electrical transport measurements for Te nanowires were performed in a Quantum Design Dynacool system with a variable temperature down to 1.8 K. Both typical four-probe and two-probe measurements showed consistent semiconducting nature of the Te crystal at room temperature. However, the large resistance due to the semiconducting gap prevents it from using the four-probe measurements at lower temperatures. We recorded reliable two-probe resistances down to lower temperatures (~ 190 K) using the standard ac lock-in technique with a low frequency (~5 Hz) voltage excitation (~ 10 mV) applied to the source electrode while recording the current at the drain while all other electrodes are floated. The electrical transport measurements for Pd-Te compound device were conducted in a dilution refrigerator with a superconducting magnet up to 8 T and a base temperature of ~ 30 mK. The resistance measurements for Pd-Te compounds were performed using the standard ac lock-in technique with a low frequency around 21 Hz and an ac current excitation of around 1 uA. A dc current was applied to the source electrode for critical current measurements.

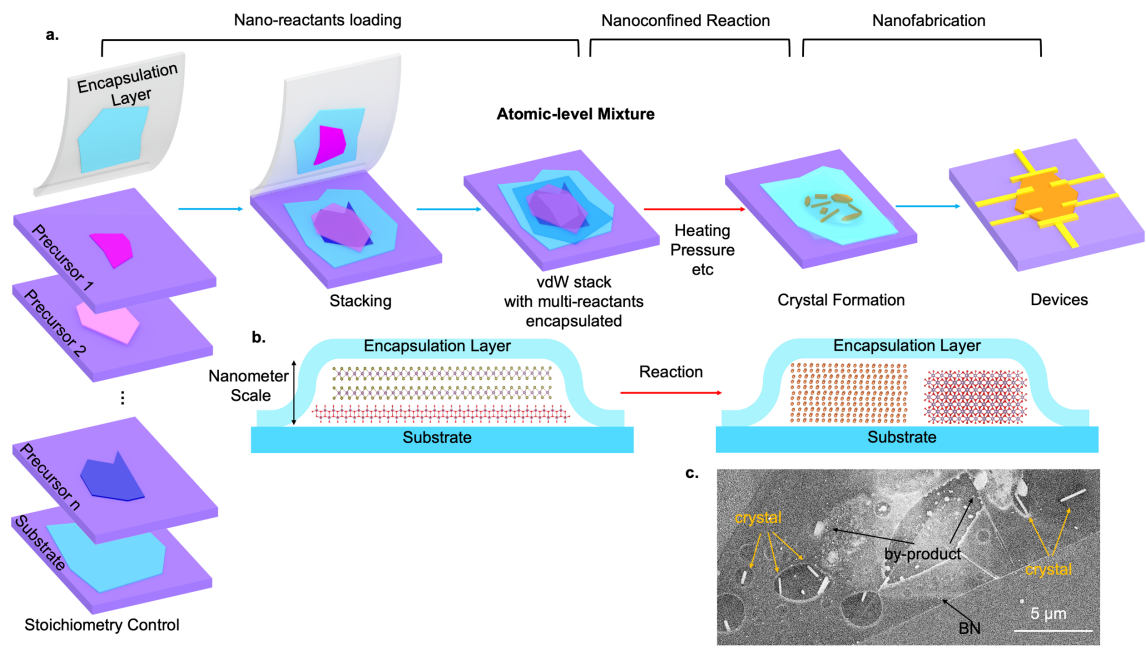

**Figure 1. van der Waals Nanoreactors. a.,** A schematic flowchart for creating vdW nanoreactors, introducing nanoconfined reactions, and fabricating nanodevices for investigating electronics properties of as-grown crystals. **b.,** A cross-sectional illustration of a representative possible nanoreactor before and after the chemical reaction. The encapsulation layer is a chemically inert vdW material, such as hBN. The substrate can be hBN or other typical rigid substrates such as $SiO_2$, sapphire etc. **c.,** An SEM image of a representative vdW nanoreactor after the reaction, showing the formation of single crystals and by-product materials. This specific vdW nanoreactor is created following a similar procedure described in Fig. 2. The crystals synthesized are high-quality tellurium nanowires, whereas the substrate is $Si/SiO_2$ and the encapsulation layer is hBN.

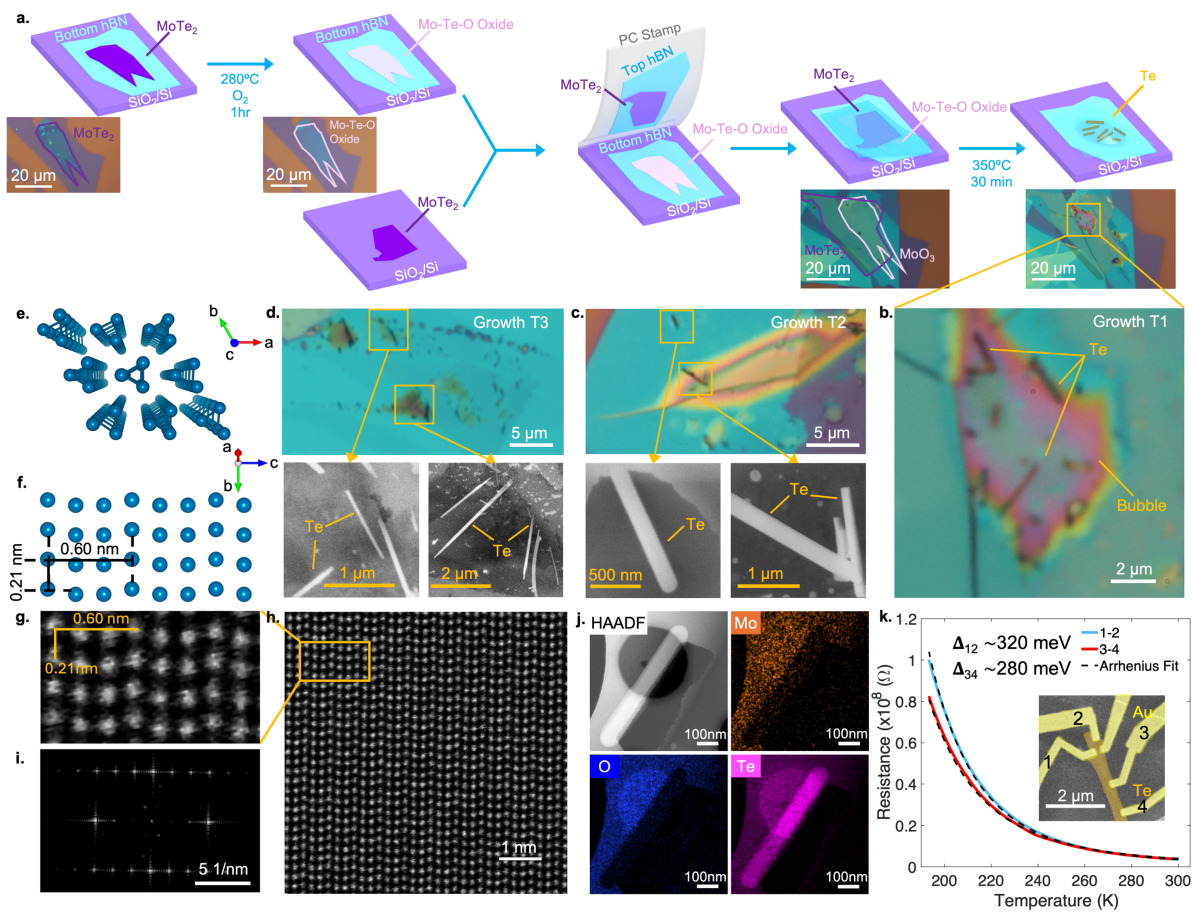

**Figure 2. vdW Nanoreactors for synthesizing tellurium single crystals. a.,** Illustration of the workflow for designing and creating the nanoreactors and the reactions. Insets are optical images of the flakes, outlined by the solid line along the boundaries, used in Growth T1 at selected steps. **b.,** An optical image of the nanoreactor after the reaction. The final products are multiple separated Te wires, some of which are indicated. The bubble is a hBN tent structure formed during the growth. **c.,** Top: The same optical image taken for another independent growth (T2) after a similar procedure. Bottom: high resolution SEM images taken at locations indicated in the top panel. **d.,** the same as **c** but for another independent growth T3, showing that the reaction is highly reproducible. **e.** & **f.,** The crystal structure of tellurium, viewed from different angles as illustrated by the compass. **g.,** An atomic resolution STEM image of a typical as-grown Te wire, showing matched crystal structure and interatomic distances. **h.,** The same STEM image but with a larger area, showing high crystallinity. Each of the Te synthesized in the reactors are single crystals. **i.,** FFT of the real space image showing the characteristic diffraction spots. **j.,** EDX analysis of a chosen wire (with an HAADF image shown at the top left panel), showing both the clean Te wire and the formation of oxides nearby. **k.,** The transport measurements of the Te wire, showing an activated behavior of the resistance measured as a function of temperature (solid lines), taken from two different pairs of contacts. The Arrhenius fit (dashed back lines) to both curves yields an activation gap as indicated. Inset is a false-colored SEM image of the device after the deposition of Au electrodes.

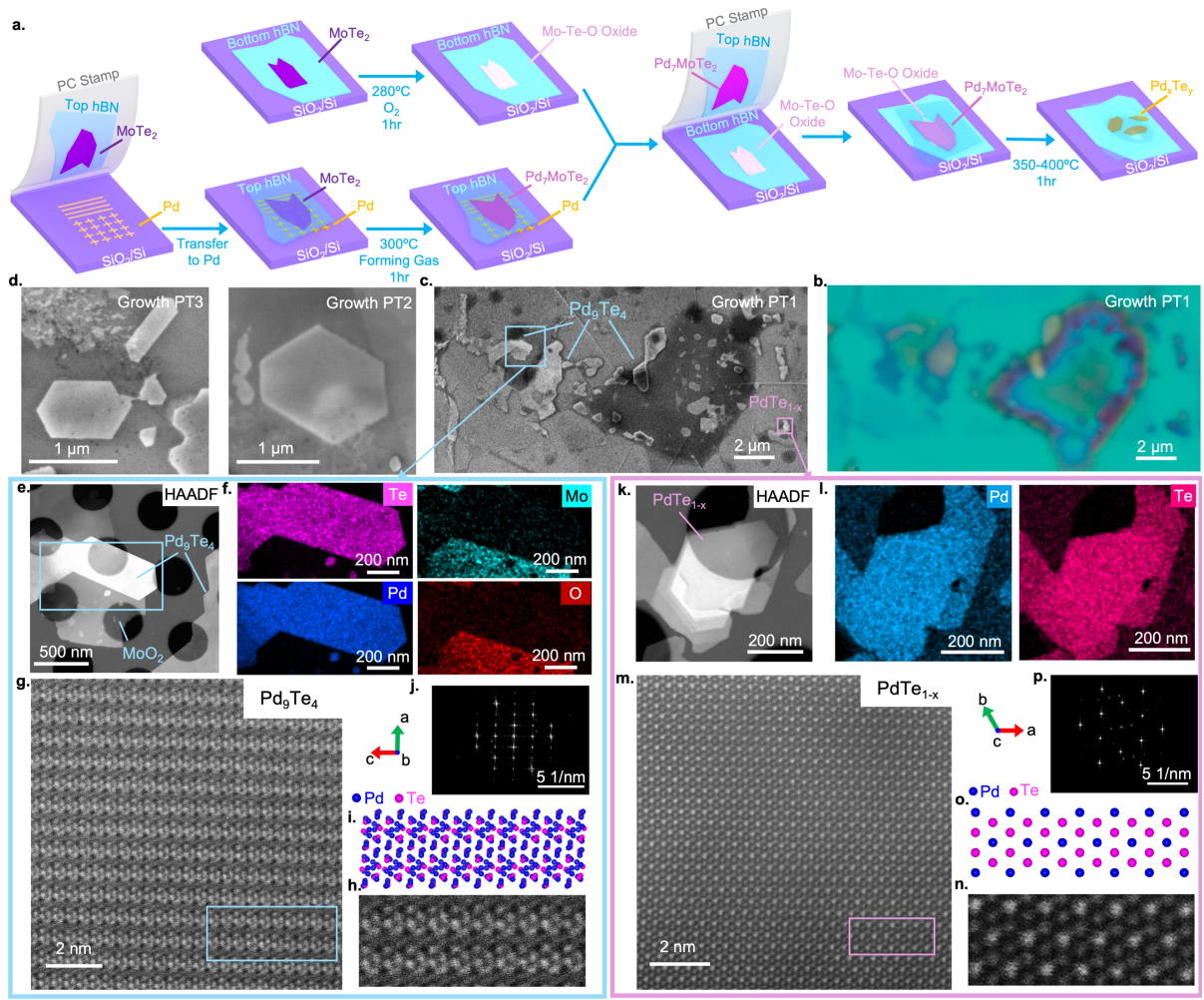

**Figure 3. van der Waals Nanoreactors for synthesizing Pd-Te crystals. a.,** Illustration of the workflow for designing and creating the corresponding nanoreactors and the reactions. **b.,** An optical image of the vdW stack after the reactions (Growth PT1). **c.,** A high-resolution SEM image of the same stack to **b**. **d.,** SEM images of two representative crystalline flakes obtained respectively from another two independently created vdW nanoreactor and reactions, following a similar procedure. **e.,** A HAADF image of a region highlighted in blue square shown in **c**. **f.,** EDX elemental mapping, with the element indicated in each panel, showing the corresponding compositions in the flakes at areas indicated by the white dashed line shown in **e**. The corresponding Pd-Te compounds are closely approximate to $Pd_9Te_4$, formed together with the nearby thin-film compound that is approximately $MoO_2$. **g.,** A representative STEM image with atomic resolution of as-grown $Pd_9Te_4$. **h.,** A zoom-in plot of the same data in the blue rectangle shown in **g**, highlighting detailed features. **i.,** The crystal structure of $Pd_9Te_4$, showing matched pattern to the data in **h**. **j.,** FFT of the real space image showing the characteristic spots of $Pd_9Te_4$. **k.,** A HAADF image of a region highlighted in purple square shown in **c**. **l.,** EDX elemental mapping, with the element indicated in each panel, showing the compositions of the flake shown in **k**. The flake consists of Pd and Te in a ratio of about 55%:45%, indicating a non-stoichiometric $PdTe_{1-x}$ with $x \approx 0.18$. **m.,** A representative STEM image with atomic resolution of as-grown $PdTe_{1-x}$. **n.,** A zoom-in plot of the same data in the purple rectangle shown in **m**. **o.,** The modeled crystal structure of $PdTe_{1-x}$, which closely resemble that of PdTe hexagonal structure with a space group $P6_3/mmc$, matching the pattern of **n**. **p.,** FFT of the real space image showing the characteristic lengths of $PdTe_{1-x}$.

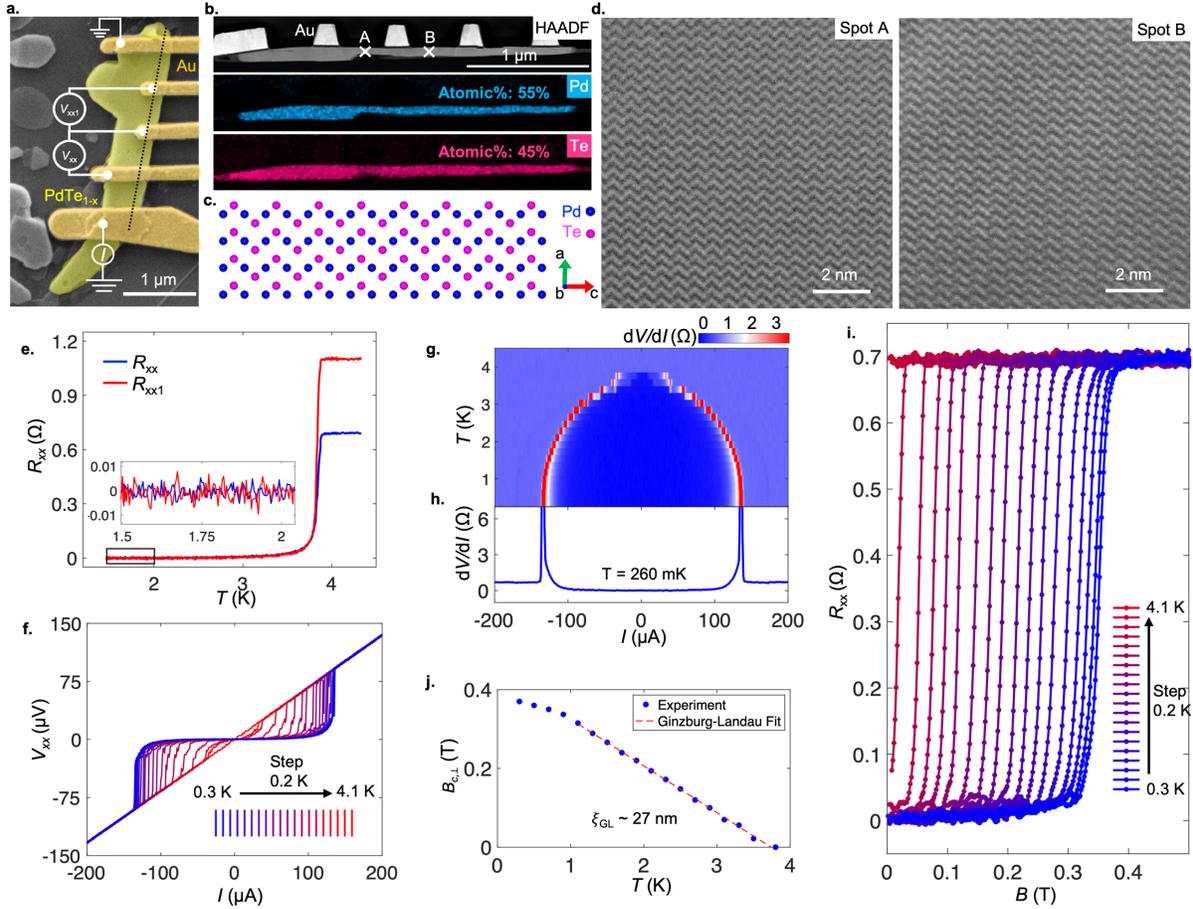

**Figure 4. Superconductivity in PdTe$_{1-x}$ (x ≈ 0.18). a.,** A false-colored SEM image of a device fabricated from as-grown PdTe$_{1-x}$ with gold (Au) electrodes deposited for electrical measurements. **b.,** Cross-section EDX analysis of the device, followed by a FIB cut along the dashed line in the same device after the transport measurement. *Top:* An HAADF image of the cross section, showing both the Au electrodes and PdTe$_{1-x}$ crystal. *Middle*: EDX elemental mapping of Pd. Bottom: elemental mapping of Te. The observed atomic ratio is indicated, which is consistent with the top view EDX mapping shown in Fig. 3l. **c.,** the modeled crystal structure (following PdTe) viewed from a different angle indicated by the compass. **d.,** Two representative STEM images with atomic resolution taken respectively at two distinct locations indicated in b (spots A and B), showing identical structures that are consistent with **c.** This data suggests that that PdTe$_{1-x}$ exhibits a highly uniform crystal structure. **e.,** Resistance of the crystal as a function of temperature. The measurement configuration for all transport data is shown in **a.** The two curves correspond to data taken from the two different contact pairs, respectively. Inset shows the zero-resistance value measured at low $T$. **f.,** IV characteristic curves taken at various $T$ indicated. **g.,** Differential resistance $dV/dI$ as a function of applied DC current ($I$), taken at different $T$, showing the $T$-dependent critical currents. **h.,** $dV/dI$ *vs* $I$ taken at $T$ = 260 mK. **i.,** $R_{xx}$ as a function of $B\perp$, taken at various $T$ as indicated. **j.,** Extracted critical magnetic field $B_{c,\perp}$ as a function of $T$. The dashed red line is the linear GL fit to the data.

**Extended Data Figures**

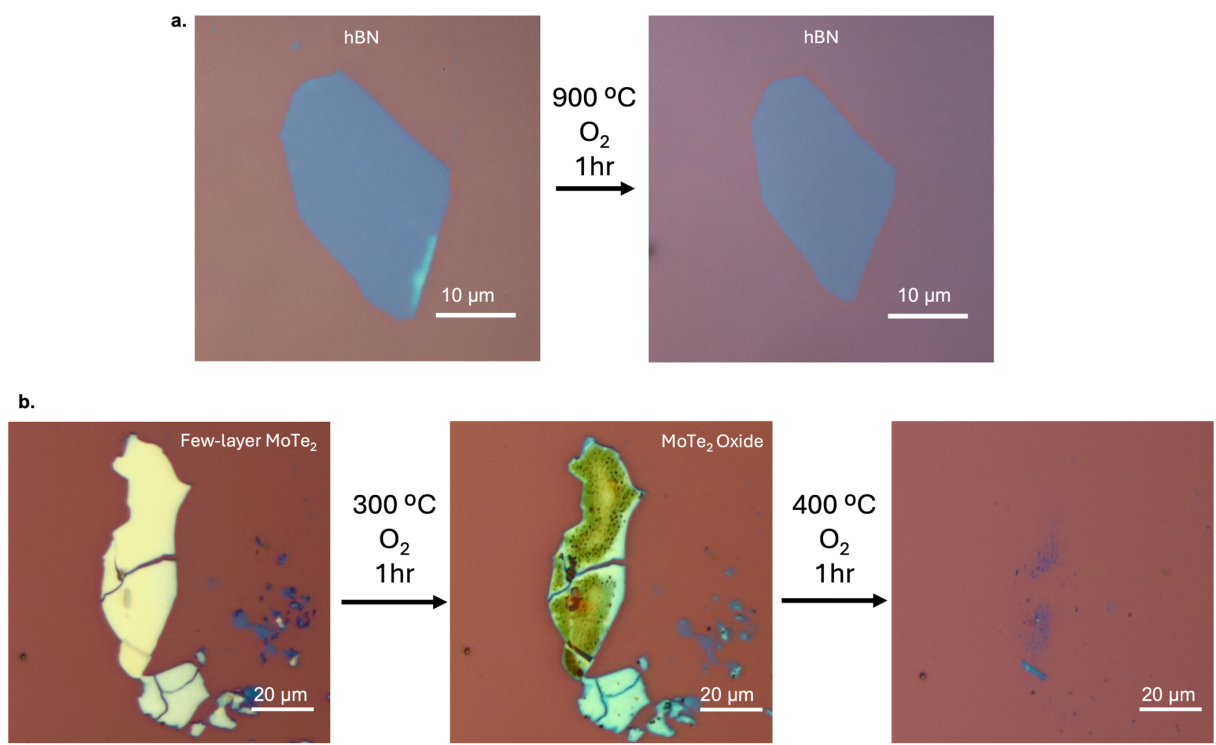

**Extended Data Fig.1 | The impact of temperature and oxidization on hBN and MoTe$_2$ a.,** Optical images of a hBN flake (15 nm thick) before and after 900 °C heat treatment in the furnace. The hBN remains stable under this high temperature, showing no evidence of oxidization. Residual glue-like contaminants at the edge are removed during the annealing process. **b.,** Optical images of exposed few-layer MoTe$_2$ (left) on SiO$_2$/Si substrate, after two successive heat treatments in an oxygen environment (middle & right). Without top hBN encapsulation, the oxidized MoTe$_2$ sublimed (right). The control experiments confirm the necessity of top encapsulation using inert materials--such as hBN—for maintaining material stability in the vdW nanoreactors demonstrated in our work.

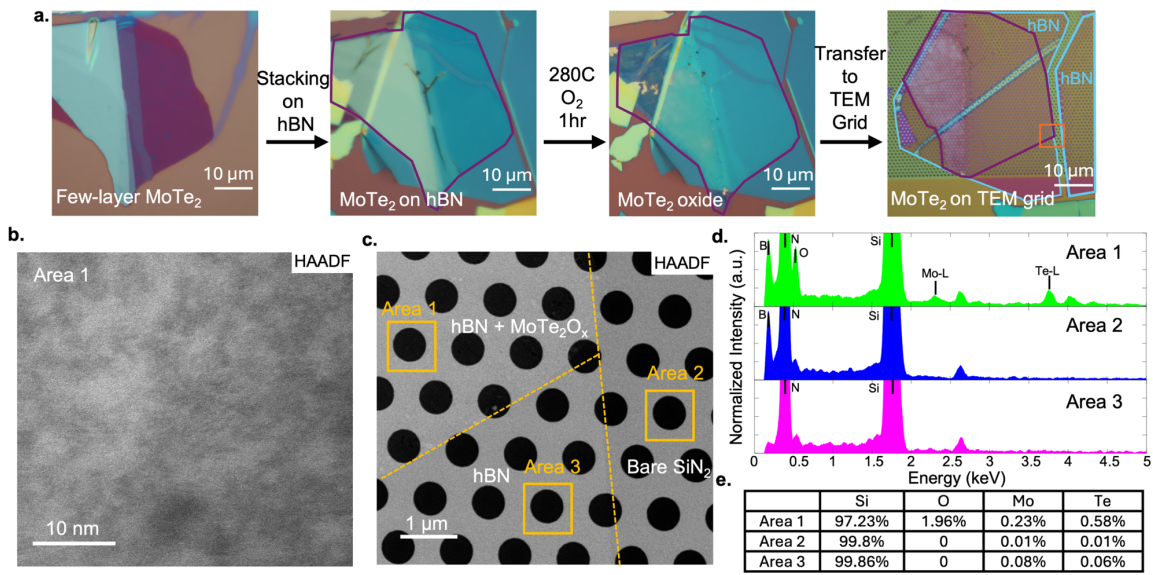

**Extended Data Fig.2 | Charaterization of Mo-Te-O Oxide. a.,** Optical images at each stage of the oxidization process of a few-layer MoTe$_2$ flake and the TEM sample preparation. Purple polygons in the last three images outline approximately the shape of MoTe$_2$ or its oxides in each case. **b.,** A typical HAADF TEM image of the oxide, taken at Area 1 shown in **c**. No obvious crystalline order is observed, indicating the formation of an amorphous structure after the oxidization treatment to the original MoTe$_2$ crystal. **c.,** An HAADF image of top view of MoTe$_2$ oxide sample, with corresponding locations indicated by the orange square in the last image in **a**. Area 1 consists of MoTe$_2$ oxide on hBN, Area 2 is bare SiN$_2$ substrate, while Area 3 is only hBN on the substrate. The dashed line indicates the phase separation. **d.,** EDX spectra of Area 1, Area 2 and Area 3 of the spots shown in **c**, respectively. The characteristic X-ray peaks corresponding to O, Mo and Te are well resolved in Area 1, while the peak corresponding to O is absent in Area 2 and 3. **e.,** Extracted atomic fractions of Si, O, Mo, and Te for Areas 1-3, respectively. Oxidized MoTe$_2$ is roughly of the composition Mo:Te:O ~ 1:2.5:8.5, consistent with the expectation of the formation of MoO$_3$ and TeO$_x$.

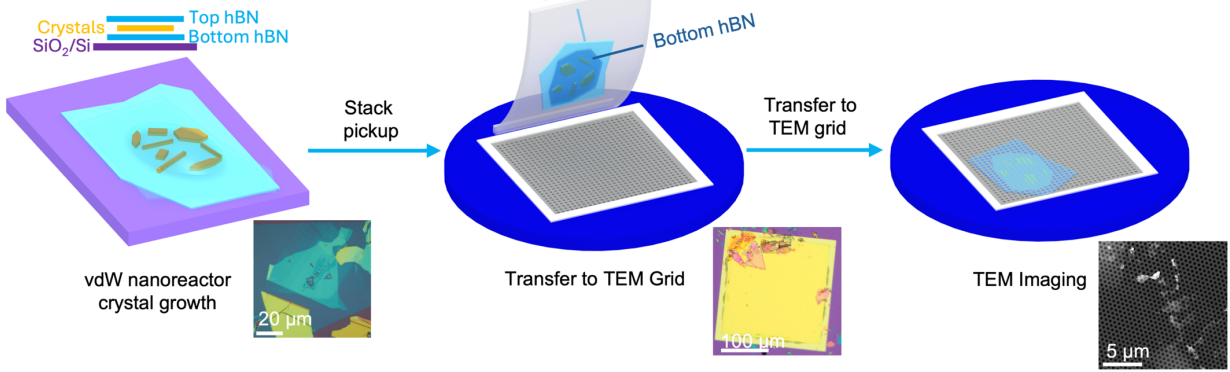
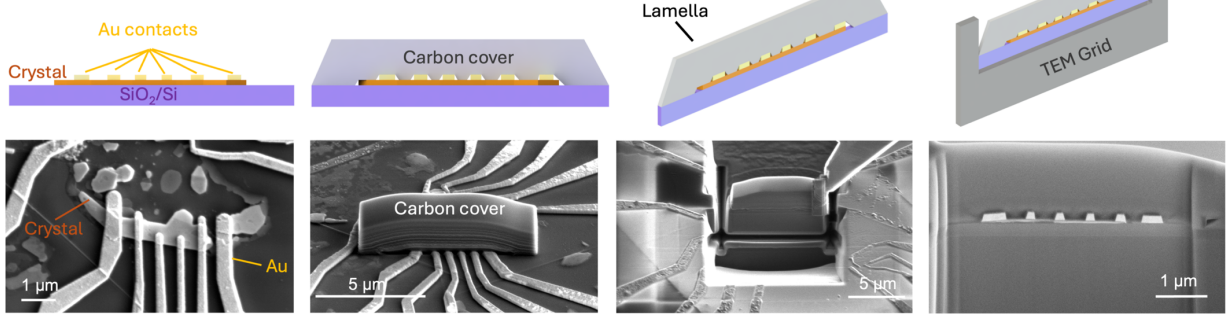

**Extended Data Fig.3 | TEM Samples. a.,** Cartoon illustrations of plan-view TEM sample preparation process. The vdW nanoreactor after final crystal growth (encapsulated by top and bottom hBN) is transferred onto holey silicon nitrate TEM grid. Insets are optical (the first two) and STEM (the last) images of a typical sample at corresponding stages. **b.,** Cartoon illustrations (top panel) and SEM images (bottom panel) of cross-section TEM sample preparation process. The transport device was first covered by a layer of carbon and a lamella was prepared using standard FIB milling process. The final lamella was transferred to a TEM grid.

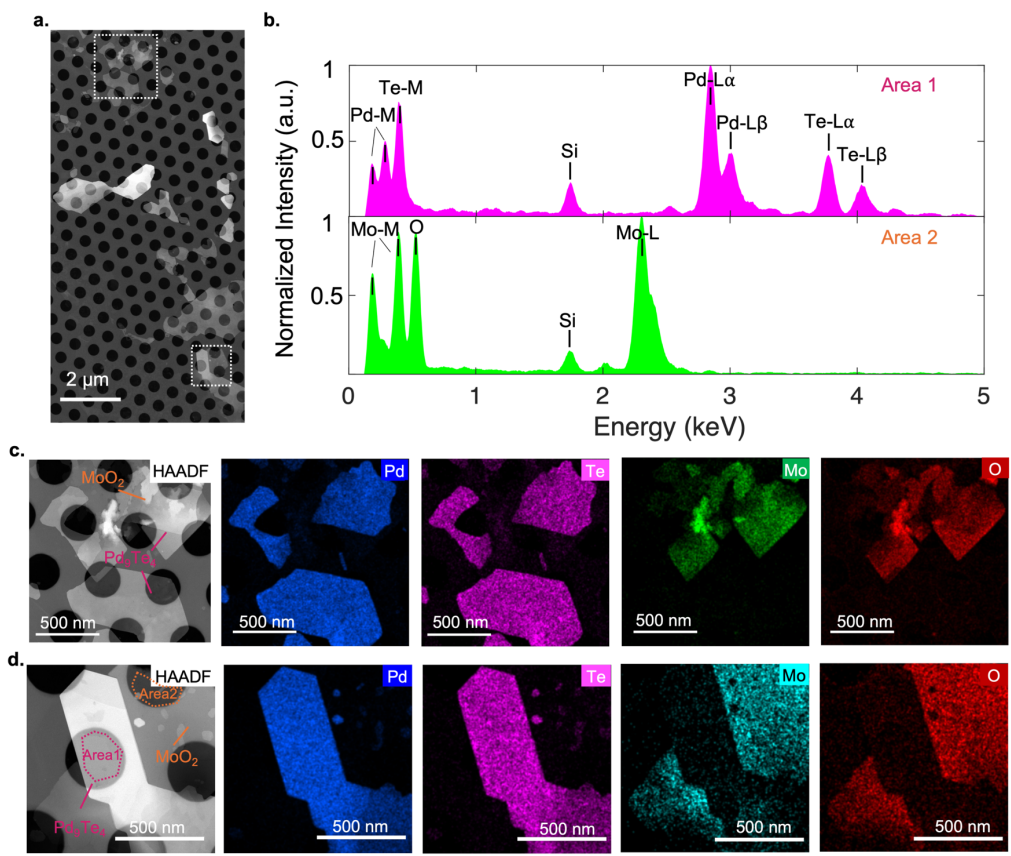

**Extended Data Fig.4 | Additional EDX Data around two Pd$_9$Te$_4$ crystals. a.,** A plan-view HAADF TEM image of the sample. **b.,** EDX spectra of Area 1 and Area 2 in the sample regions shown in **d**, which zooms into the bottom white square in **a**. The characteristic X-ray energy peaks corresponding to electron relaxation to the L/M shell of Pd and Te, namely, Pd-M, Te-M, Pd-L and Te-L, are well resolved Area 1, while Mo-L and O peaks emerge only in Area 2. **c & d.,** EDX elemental mapping of Pd, Te, Mo, and O corresponding to the samples in two boxed areas (by white dashed lines) in **a**, respectively. The corresponding Pd-Te compounds are closely approximate to Pd$_9$Te$_4$, formed together with the nearby thin-film compound that is approximately MoO$_2$.

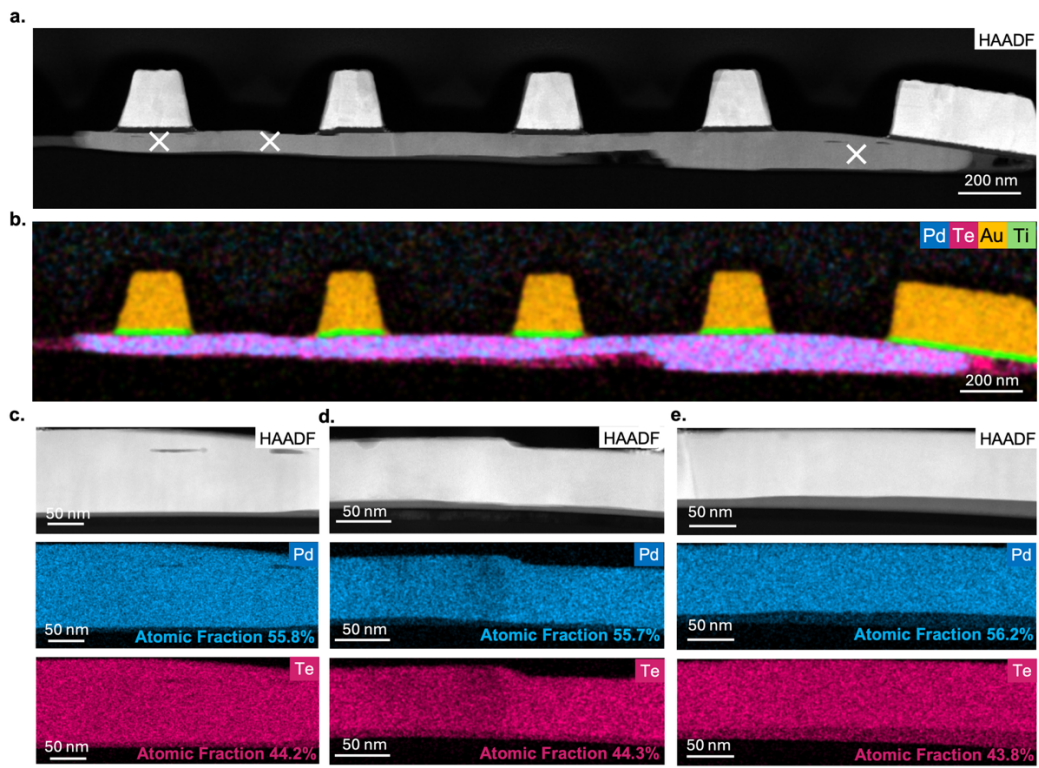

**Extended Data Fig.5 | Additional EDX Data on PdTe$_{1-x}$ a.,** HAADF TEM image of cross-section view on PdTe$_{1-x}$ with deposited Au electrodes. **b.,** EDX elemental mapping of all elements detected, including Au and Ti from electrode deposition and Pd and Te from the as-grown crystal. **c-e.,** Cross-sectional EDX analysis of the slice at three different locations as indicated by the white cross in **a,** highlighting the mapping of Pd and Te for the crystal. The observed atomic ratio is indicated in each mapping, suggesting a high uniformity across the device.

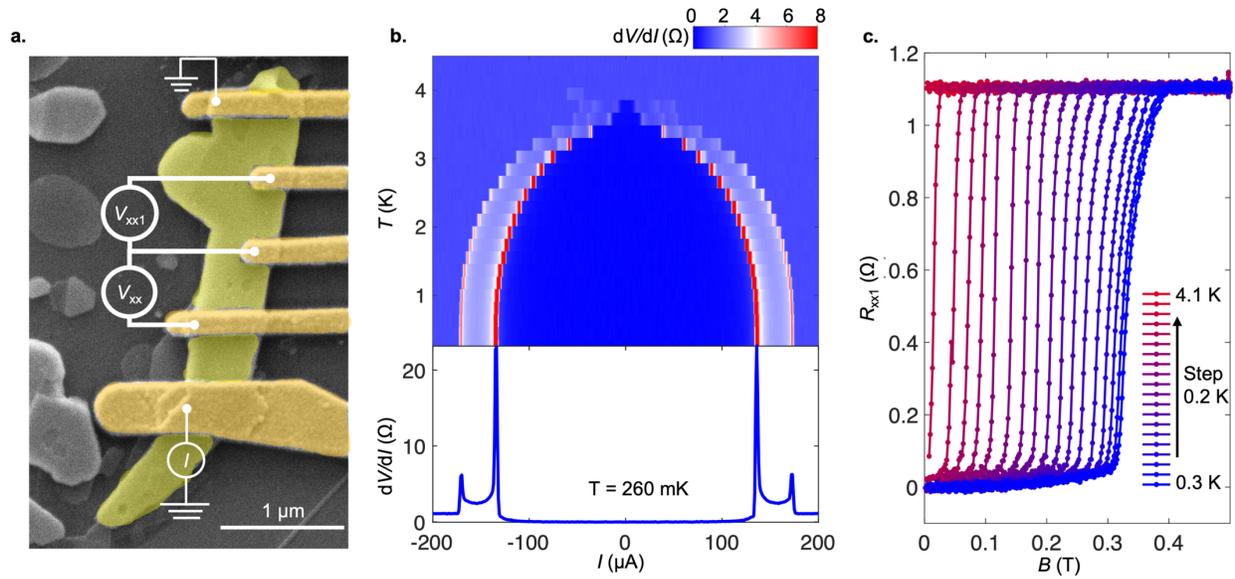

**Extended Data Fig.6 | Additional Transport Measurement for PdTe$_{1-x}$ a.,** A false-colored SEM image of a device made of as-grown PdTe$_{1-x}$ with deposited Au electrodes. **b.,** Differential resistance d$V$/d$I$ as a function of applied DC current ($I$), taken at different $T$, showing the $T$-dependent critical currents measured from Vxx$_1$, as illustrated in **a**. The bottom panel is a single curve d$V$/d$I$ vs $I$ taken at $T$ = 260 mK. **c.,** $R_{xx1}$ as a function of $B\perp$, taken at various $T$ as indicated by the color.